\documentclass[12pt]{article}
\usepackage{a4wide}
\usepackage{amssymb, amsmath}
\usepackage[normalem]{ulem}
\usepackage{graphicx}
\usepackage{subfigure}
\usepackage{cite}
\usepackage[utf8]{inputenc}
\usepackage[colorlinks]{hyperref}
\usepackage[usenames,dvipsnames]{color}
\hypersetup{
     breaklinks=true,
    pdfstartview={FitH},    
    colorlinks=true,       
    linkcolor=blue,          
    citecolor=red,        
    filecolor=magenta,      
    urlcolor=blue,           
    anchorcolor=green,      
    linktocpage=true
}

\begin{document}
{\renewcommand{\thefootnote}{\fnsymbol{footnote}}
		
\begin{center}
{\LARGE Relating the scalar weak gravity conjecture and the swampland distance conjecture for an accelerating universe} 
\vspace{1.5em}

Suddhasattwa Brahma$^{1}$\footnote{e-mail address: {\tt suddhasattwa.brahma@gmail.com}} and Md. Wali Hossain$^{1}$\footnote{e-mail address: {\tt wali.hossain@apctp.org}}
\\
\vspace{0.5em}
$^1$ Asia Pacific Center for Theoretical Physics, Pohang 37673, Korea\\
\vspace{1.5em}
\end{center}
}
	
\setcounter{footnote}{0}

\newcommand{\bea}{\begin{eqnarray}}
\newcommand{\eea}{\end{eqnarray}}
\renewcommand{\d}{{\mathrm{d}}}
\renewcommand{\[}{\left[}
\renewcommand{\]}{\right]}
\renewcommand{\(}{\left(}
\renewcommand{\)}{\right)}
\newcommand{\nn}{\nonumber}
\newcommand{\Mpl}{M_{\textrm{Pl}}}
\def\H{\mathrm{H}}
\def\V{\mathrm{V}}
\def\e{\mathrm{e}}
\def\be{\begin{equation}}
\def\ee{\end{equation}}

\def\al{\alpha}
\def\bet{\beta}
\def\gam{\gamma}
\def\om{\omega}
\def\Om{\Omega}
\def\sig{\sigma}
\def\Lam{\Lambda}
\def\lam{\lambda}
\def\ep{\epsilon}
\def\ups{\upsilon}
\def\vep{\varepsilon}
\def\S{\mathcal{S}}
\def\doi{http://doi.org}
\def\arxiv{http://arxiv.org/abs}
\def\d{\mathrm{d}}
\def\g{\mathrm{g}}
\def\m{\mathrm{m}}
\def\r{\mathrm{r}}

\begin{abstract}
\noindent For any quasi de Sitter background, we show that a recently proposed scalar weak gravity conjecture (sWGC) follows from the swampland distance conjecture, together with the covariant entropy bound. While pointing out the limitations of our argument, we suggest how further generalizations of our result might indicate that the shape of the potential of a scalar field in a low-energy effective field theory gets highly constrained due to the distance conjecture alone, going beyond the refined de Sitter conjecture. We also correct some previous comments regarding cosmological implications of the sWGC.
\end{abstract}

\section{Introduction}
There has been a lot of emphasis in recent times on how fundamental descriptions of nature can be brought closer to observable phenomenology. One of the most promising developments has been how aspects of ultraviolet (UV) physics can show up in model-building for describing phenomena in the infrared (IR). Specifically, it has been postulated that effective field theories (EFTs) can be severely constrained when requirements of a quantum-gravity completion are taken into account \cite{Vafa:2005ui}. For instance, string theory constructions seem to suggest restrictions on the allowed shape of potentials for a scalar field, even when viewed as a low energy EFT coupled to gravity \cite{Obied:2018sgi}. If true, these type of requirements point towards a very powerful test of consistency for low-energy models with quantum gravity. The situation, it seems, is qualitatively similar to early days of quantum theory when, say, the angular momentum of an electron was found to take only certain values, as opposed to the unrestricted range allowed by classical physics. Interestingly, almost all of these consistency conditions, the so-called `swampland conjectures' (see, \textit{e.g.} \cite{Brennan:2017rbf, Palti:2019pca} and references therein), come in the form of inequalities, reminiscent of the seminal role played by the uncertainty principle in the development of quantum mechanics. These conjectures have their footing in strong logical reasoning and most of them have been tested extensively in string theory constructions and have, thus far, been found to hold true.

The natural next step would be to prove these conjectures from first principles. However, this seems to be a very daunting task as it is and a more tractable problem would be to check the consistency of (some of) these conjectures, assuming one or more of the others to hold. Indeed, this was shown to be the case with the so-called `de Sitter (dS) swampland conjecture'. Albeit it was initially motivated by the difficulty of constructing meta-stable dS solutions in string theory, it was shown to be true \cite{Ooguri:2018wrx}, given the `swampland distance conjecture' \cite{Ooguri:2006in,Baume:2016psm,Klaewer:2016kiy,Grimm:2018cpv,Blumenhagen:2018nts,Landete:2018kqf} and the covariant entropy bound \cite{Bousso:1999xy,Fischler:1998st}. We follow a similar philosophy in this work to investigate a recently proposed scalar weak gravity conjecture (sWGC) \cite{Gonzalo:2019gjp}. Although the authors of \cite{Gonzalo:2019gjp} had proposed this sWGC from a completely different chain of reasoning, we shall show that it can indeed also be motivated given the swampland distance conjecture, at least for an accelerating universe with a quasi-dS background. We emphasize that our result goes beyond showing the compatibility of the distance conjecture and the sWGC; rather, we arrive at the latter conjecture only by assuming the former and the covariant entropy bound.

We shall conclude with a brief discussion on the phenomenological implications of the sWGC for inflation. In this section, we shall correct a mistake in the original paper \cite{Gonzalo:2019gjp}\footnote{While we were preparing this draft, the oversight was corrected in a subsequent version. However, our main argument remains unaffected as can be seen in the `note added' (Sec.~\ref{Note}) at the end.} and show that this conjecture is far less constraining for inflationary models than what was initially made out to be. We also point out the main limitations of our argument and go on to hypothesize further extensions of our work, specifically how the shape of the potential might possibly be subjected to an infinite number of constraints arising from of the distance conjecture. 

\section{The Swampland conjectures}
The Swampland conjectures are aimed to categorize apparently consistent-looking low energy EFTs which, when coupled to gravity, cannot be completed in the UV. In other words, these conjectures need to be satisfied by any EFT consistent with quantum gravity. Since string theory is believed to lead to a vast `landscape' of low-energy EFTs, this implies an even larger `swampland' of inconsistent EFTs surrounding it (see \cite{Palti:2019pca} for a review).

Recently, keeping in mind the difficulty of constructing (meta-)stable dS vacua in string theory with controlled approximations, it was proposed that the gradient of the potential $V(\phi)$ for any scalar field must satisfy the lower bound (or, must have very large tachyonic instabilities in the potential) \cite{Obied:2018sgi,Andriot:2018ept,Garg:2018reu}
\begin{eqnarray}\label{SwampdS}
	\frac{|V'(\phi)|}{V(\phi)} \ge \frac{c}{\Mpl}\;\;\;\;\;\;\;\;\; \text{or}\;\;\;\;\;\;\;\;\; \frac{V''(\phi)}{V(\phi)} \le -\frac{\tilde{c}}{\Mpl^2}\,,
\end{eqnarray}
where $c, \tilde{c} > 0$ are $\mathcal{O}(1)$ constants. This first bound, inspired by stringy constructions, excludes (meta-)stable dS vacua while the refined version allows for local maxima with large curvatures while still ruling out meta-stable dS. This `dS swampland' conjecture \eqref{SwampdS} has led to severely constraining both models of inflation \cite{Agrawal:2018own,Kinney:2018nny,Riotto}\footnote{However, see \cite{Brahma:2018hrd,Das:2018hqy,Das:2018rpg,Ashoorioon:2018sqb,Lin:2018kjm,Dimopoulos:2018upl,Motaharfar:2018zyb,Kinney:2018kew} for proposals on how to evade these bounds for single-field inflation.} as well as late-time acceleration \cite{Raveri:2018ddi,Heisenberg:2018yae,Heisenberg:2019qxz,Brahma:2019kch,Odintsov:2018zai,Elizalde:2018dvw,Heckman:2019dsj,Heckman:2018mxl,Mukhopadhyay:2019cai}. In fact, current observational data has also been used to estimate how large the value of these $\mathcal{O}(1)$ constants can be: \textit{e.g.}, it was constrained that $c<0.6$ from present acceleration of the universe \cite{Heisenberg:2018yae}.  

Although this `dS conjecture' was originally primarily motivated by string theory examples, it was shown in \cite{Ooguri:2018wrx} that it follows from more general arguments based on the covariant entropy bound for (a causal patch of) dS space and the swampland distance conjecture. The general message of this latter conjecture is that an EFT is valid only for a finite field variation due to the appearance of an infinite tower of states which become exponentially light \cite{Ooguri:2006in}. The more refined version of this conjecture \cite{Palti:2019pca,Blumenhagen:2018nts} states the following: Given the moduli space of string vacua parametrized by the expectation value of some fields $\phi^i$ which have no potential, for two points on this moduli space separated by a large geodesic distance $\Delta \phi > \Mpl$, there appears an infinite tower of light states with masses $m \sim e^{-a\Delta\phi/\Mpl}$ for some $\mathcal{O}(1)$ parameter $a>0$. The kinetic term of the scalar fields provide a metric for measuring distances in field space. This conjecture has not only been extensively tested in string theory (see, for instance, \cite{Grimm:2018ohb,Lee:2018urn}), but also generalized in the presence of scalar potentials \cite{Baume:2016psm,Klaewer:2016kiy}. Without going into the details of these statements, the widely accepted conclusion of the swampland distance conjecture has been that in any EFT, the field excursion $\Delta\phi$  has an upper limit and that it cannot be super-Planckian \cite{Palti:2019pca}. 

Since the `dS conjecture' seemed to be somewhat radical at first \cite{Akrami:2018ylq,Kallosh:2019axr}, especially due to its implications for inflation, it was reassuring to find that it can be shown to follow logically from the distance conjecture, coupled with Bousso's covariant entropy bound for quasi-dS backgrounds \cite{Ooguri:2018wrx}. Indeed, applying this line of reasoning also neutralized an array of counter-examples \cite{Denef:2018etk,Conlon:2018eyr,Choi:2018rze,Murayama:2018lie,Hamaguchi:2018vtv}, which were proposed to invalidate the original `dS conjecture', through refining it and, in the process, avoiding its application to local maxima in the potential. 

\section{The sWGC from the distance conjecture}
The general principle that gravity is the weakest force has led to another set of conjectures, collectively known as the Weak Gravity Conjecture (WGC) \cite{Vafa:2005ui,ArkaniHamed:2006dz}\footnote{Related motivations for the WGC comes from the perspective of stability of extremal charged black-holes and from various stringy examples (see \textit{e.g.} \cite{Hamada:2018dde, Harlow:2015lma, Heidenreich:2016aqi, Montero:2016tif, Cheung:2018cwt, Andriolo:2018lvp} for some recent literature on this).}. Very recently, it has also been conjectured \cite{Gonzalo:2019gjp} that for any scalar field, coupled to quantum gravity, the potential of the field must satisfy some constraints such that self-interactions of the scalar field do not become weaker than gravity. This statement, in the case of a single (canonically normalized) scalar field with potential $V(\phi)$, is quantified as follows\footnote{This was how the sWGC was originally formulated in \cite{Gonzalo:2019gjp}. As mentioned before, during the final stages of our draft, we noticed that the explicit form of the sWGC was modified in subsequent revision of \cite{Gonzalo:2019gjp}. We checked to ensure that our arguments in this section remain completely unaffected by this alteration (see our `note added' (Sec.~\ref{Note}) for further clarifications).}
\begin{eqnarray}\label{SWGC}
	2 \frac{(V''')^2}{V''} - V'''' -\frac{V''}{\Mpl^2} > 0 \,.
\end{eqnarray}
As has already been discussed in \cite{Gonzalo:2019gjp}, this conjecture is consistent with expectations from the swampland distance conjecture \cite{Ooguri:2006in,Blumenhagen:2018nts}. In what follows, we shall try to go in the other direction in the spirit of \cite{Ooguri:2018wrx}. Assuming the swampland distance conjecture, we shall motivate that the sWGC, as defined above, follows naturally when combined with the covariant entropy bound \cite{Bousso:1999xy} for an accelerating universe.

Let us assume that we have an accelerating background, quantified by the relation $|V'|\leq \sqrt{2}\, V$ (in terms of the so-called slow-roll parameter, this implies $\epsilon \le 1$). From now on, we shall work in Planck units and set $\Mpl =1$. At parametrically large distances in field space, the distance conjecture implies that towers of states become exponentially light leading to monotonically increasing the number of states, and consequently, the dimension of the Hilbert space. Therefore, if we associate an entropy with an observer's causal patch on a quasi-dS background, then this entropy would increase monotonically with the growing field range. On the other hand, there is a covariant entropy bound which can be applied to the largest possible area -- the apparent horizon -- of the quasi-dS background. As long as the apparent horizon of the accelerating background remains within the event horizon defined by the late time value of the potential as the field rolls down it, we are allowed to apply the covariant entropy bound to this area. The important step in our procedure is to assume that the entropy of the states in the causal patch of the accelerating background, when the Hilbert space is dominated by the emerging tower of light states, would saturate this entropy bound. It is precisely this assumption which allows us to estimate the behaviour of the potential in terms of the number of light states in the tower. We flesh out the details of this argument in the rest of this section.

More concretely, following \cite{Ooguri:2018wrx}, let us assume that a natural measure of the entropy associated with a quasi-dS background can be given by $S_{\rm qdS}\sim R^2 \sim 1/V$, where $R$ is the radius of the apparent horizon. This is done in analogy with the Gibbons-Hawking entropy for dS space \cite{Gibbons:1977mu}, in which case $V =\Lambda$ is a constant. As the field traverses trans-Planckian distances, one expects towers of light states to descend from the UV as predicted by the distance conjecture. This leads to an increase in the entropy associated with these towers of states. Although it is not established how one can calculate this entropy from a precise counting of microstates, it is still possible to estimate that the entropy would scale as 
\begin{eqnarray}
	S_{\rm tower}\sim N^\gamma R^\delta\,,
\end{eqnarray}
where $\gamma, \delta$ are some $\mathcal{O}(1)$ numbers. For an accelerating universe with an apparent horizon, one can apply the covariant entropy bound \cite{Bousso:1999xy} to get 
\begin{eqnarray}
	S_{\rm tower} \le S_{\rm qdS}\,.
\end{eqnarray}
When the tower dominates the Hilbert space, \textit{i.e.} for $\delta<2$, $N$ increases exponentially and $R$ changes to compensate adequately  so as to not have a violation of the bound. In this limit, the covariant entropy bound shall get saturated and we get the relation
\begin{eqnarray}\label{Pot}
	V(\phi) \sim R^{-2} \sim N^{-2\gamma/(2-\delta)}\,.
\end{eqnarray}
Saturating the bound implies that our derivation is valid only for paramtrically large distances in field space, just as is the case for \cite{Ooguri:2018wrx}. The number of light states in the tower is parametrized as \cite{Ooguri:2018wrx}
\begin{eqnarray}\label{N}
	N(\phi) \sim n(\phi)\, e^{\lambda\phi}\,,
\end{eqnarray}
where $n(\phi)$ signifies the effective number of different towers becoming exponentially light, and is expected to increase monotonically with $\phi$, and $\lambda$ is some $\mathcal{O}(1)$ number.

We define $c = 2\lambda\gamma/(2-\delta)$, another $\mathcal{O}(1)$ parameter, for future convenience\footnote{This is precisely the $\mathcal{O}(1)$ number which appears in the dS conjecture, $|V'|\ge c V$.}. Given \eqref{Pot} and \eqref{N}, one can write down the derivatives of the potential as follows (where $'$ denotes a derivative with respect to $\phi$)
\begin{eqnarray}
	V'   &=& - c \left[1+\frac{n'}{n \lambda}\right] V \, ,\label{V1}\\
	V''  &=&  c\left[c+\frac{2c}{\lam}\frac{n'}{n}+\frac{1}{\lam}\(1+\frac{c}{\lam}\)\(\frac{n'}{n}\)^2-\frac{1}{\lam}\frac{n''}{n}\right] V  \, , \label{V2}\\
	V''' &=& -c\left[c^2+\frac{3}{\lam}\left\{c^2-\(1+\frac{c}{\lam}\)\frac{n''}{n}\right\}\frac{n'}{n}+\frac{3c}{\lam}\(1+\frac{c}{\lam}\)\(\frac{n'}{n}\)^2 \right. \nn \\ && \left. +\frac{1}{\lam}\(2+\frac{3c}{\lam}+\frac{c^2}{\lam^2}\)\(\frac{n'}{n}\)^3-\frac{3c}{\lam}\frac{n''}{n}+\frac{1}{\lam}\frac{n'''}{n}\right]V \, , \label{V3}\\
	V'''' &=&  c\Bigg[c^4+\frac{4c}{\lam}\left\{c^3-\(1+\frac{c}{\lam}\)\(3c\frac{n ''}{n}-\frac{n'''}{n}\)\right\}\frac{n'}{n}+\frac{6c}{\lam}\Bigg\{c^2\(1+\frac{c}{\lam}\) \nn \\ && -\(2+\frac{3c}{\lam}+\frac{c^2}{\lam^2}\)\frac{n''}{n}\Bigg\}\(\frac{n'}{n}\)^2+\frac{4c^2}{\lam}\(2+\frac{3c}{\lam}+\frac{c^2}{\lam^2}\)\(\frac{n'}{n}\)^3 \nn \\ && +\frac{c}{\lam}\(6+\frac{11c}{\lam}+\frac{6c^2}{\lam^2}+\frac{c^3}{\lam^3}\)\(\frac{n'}{n}\)^4-\frac{6c^3}{\lam}\frac{n''}{n}+\frac{3c}{\lam}\(1+\frac{c}{\lam}\)\(\frac{n''}{n}\)^2 \nn \\ && +\frac{4c^2}{\lam}\frac{n'''}{n}-\frac{c}{\lam}\frac{n''''}{n}\Bigg] V\,. \label{V4}
\end{eqnarray}
Given these expressions, we can define a quantity $\chi(\phi)$ as follows
\begin{eqnarray}\label{Chi}
	\chi(\phi):= 2 \frac{\left(V'''(\phi)\right)^2}{V''(\phi)} - V''''(\phi) - V''(\phi)\,,
\end{eqnarray}
and check to see if this quantity is always greater than zero, as claimed by the recent sWGC. So far, we did not assume any specific form for $n(\phi)$ but in order to proceed with our argument below, we shall need to make some assumptions on the behaviour of $n(\phi)$ -- the number of light towers emerging in this limit -- beyond the natural assumption of monotonically increasing as was done in \cite{Ooguri:2018wrx}. Here, we shall further assume that it has a power law behaviour, parametrized as $n(\phi) \sim \phi^\al$, where $\al \ge 1$ is a constant. 
Although such a polynomial dependence is chosen arbitrarily at this point, we note that this has been done only to concretely quantify the field dependence; however, the reliance of our conclusions on this assumption shall be critically evaluated later and it would be shown that violations of the sWGC would require functional dependence of $n$ with inflection points or concave shapes with large curvatures and other such pathologies. We discuss more about this in the last section.

\begin{figure}[t]
	\centering
	\subfigure[$\lam=0.1$ and $\al=6$.]{\includegraphics[scale=.55]{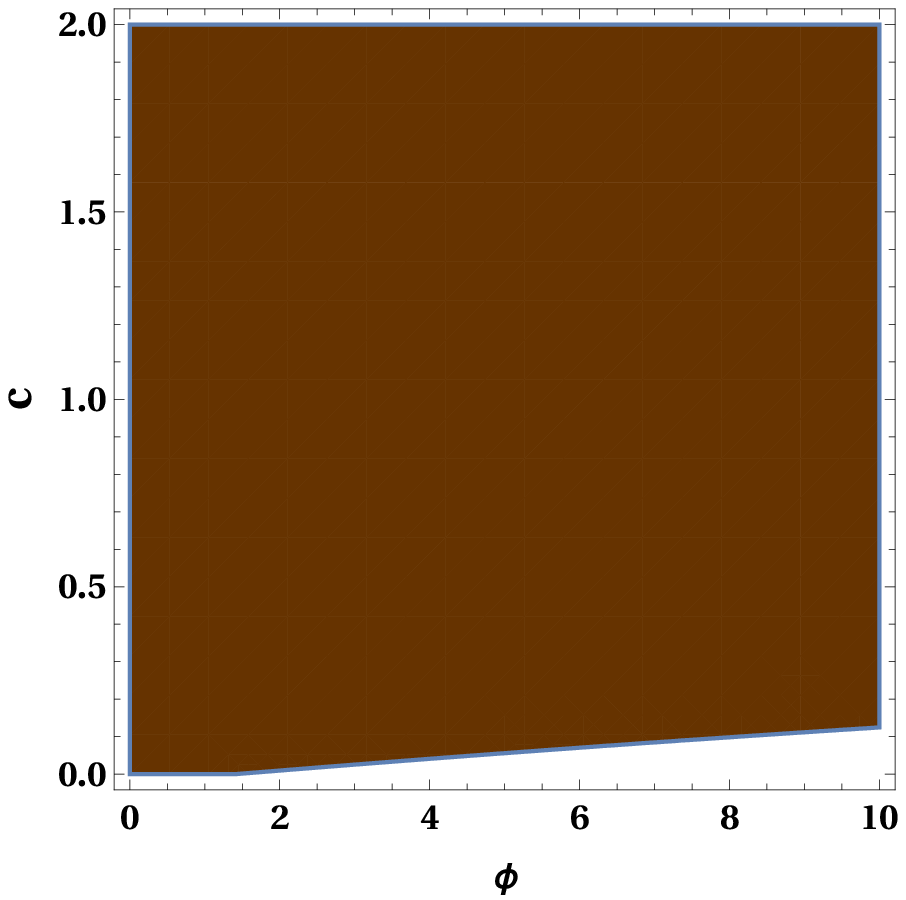}\label{fig:m6lam01}}~~
	\subfigure[$\lam=1$ and $\al=6$.]{\includegraphics[scale=.55]{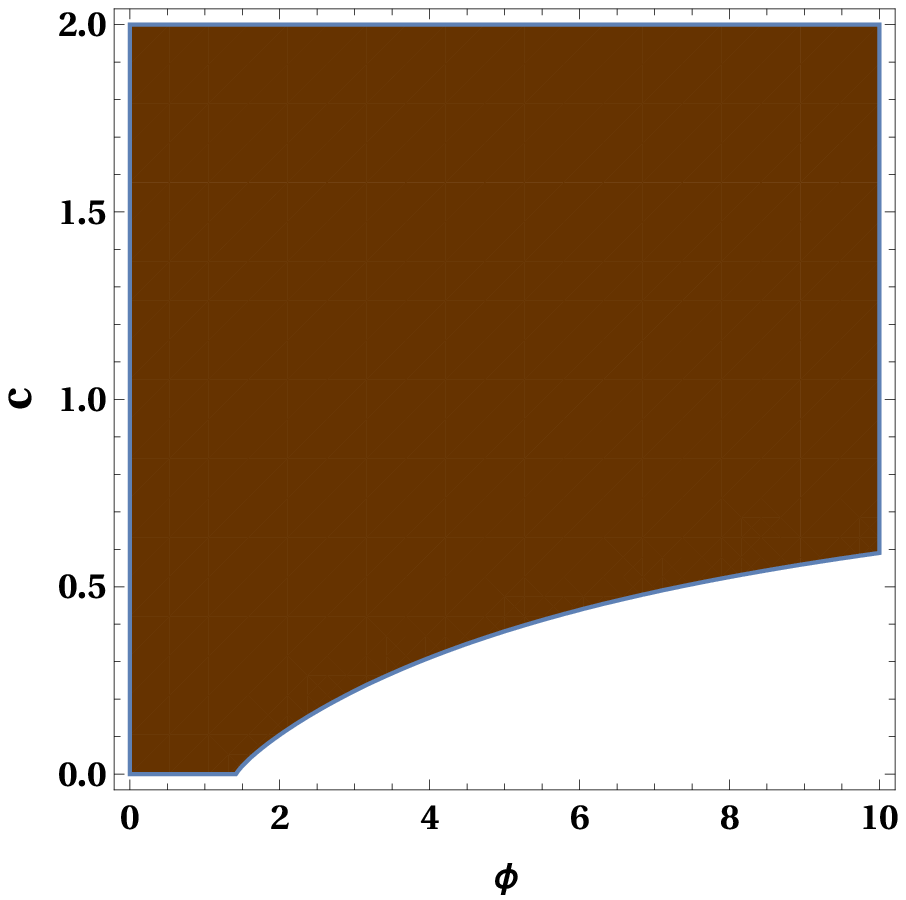}\label{fig:m6lam1}}~~
	\subfigure[$\lam=2$ and $\al=6$.]{\includegraphics[scale=.55]{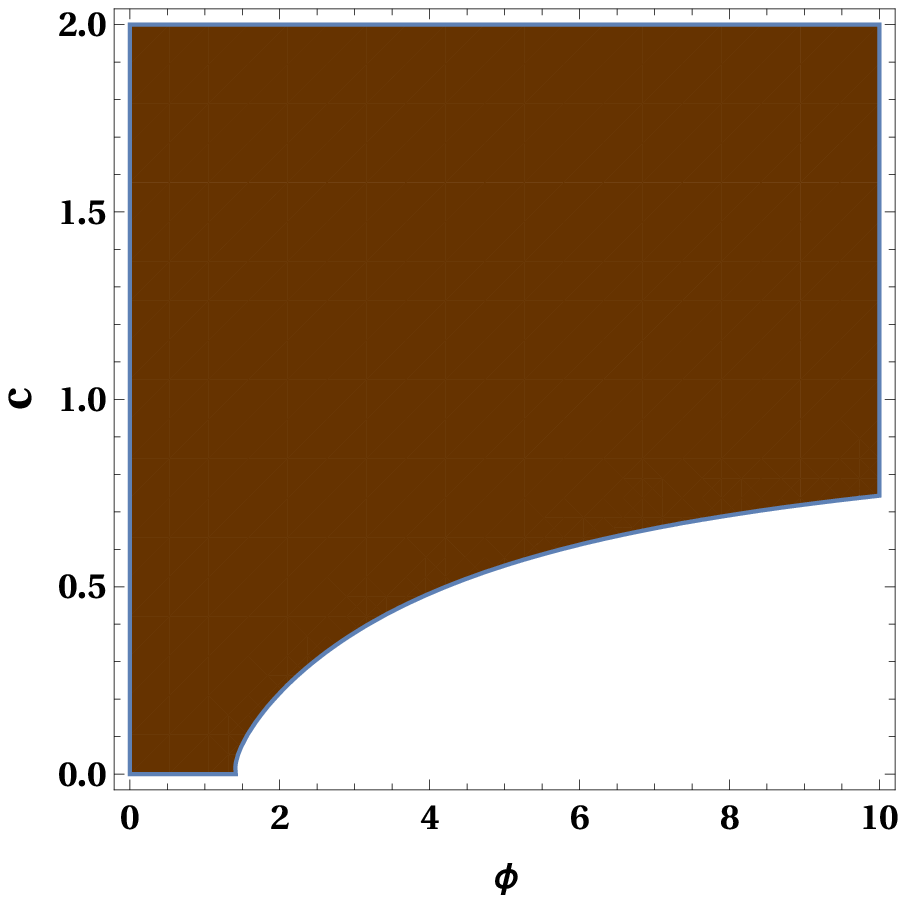}\label{fig:m6lam2}}\\
	\subfigure[$\al=1$ and $\lam=1$.]{\includegraphics[scale=.55]{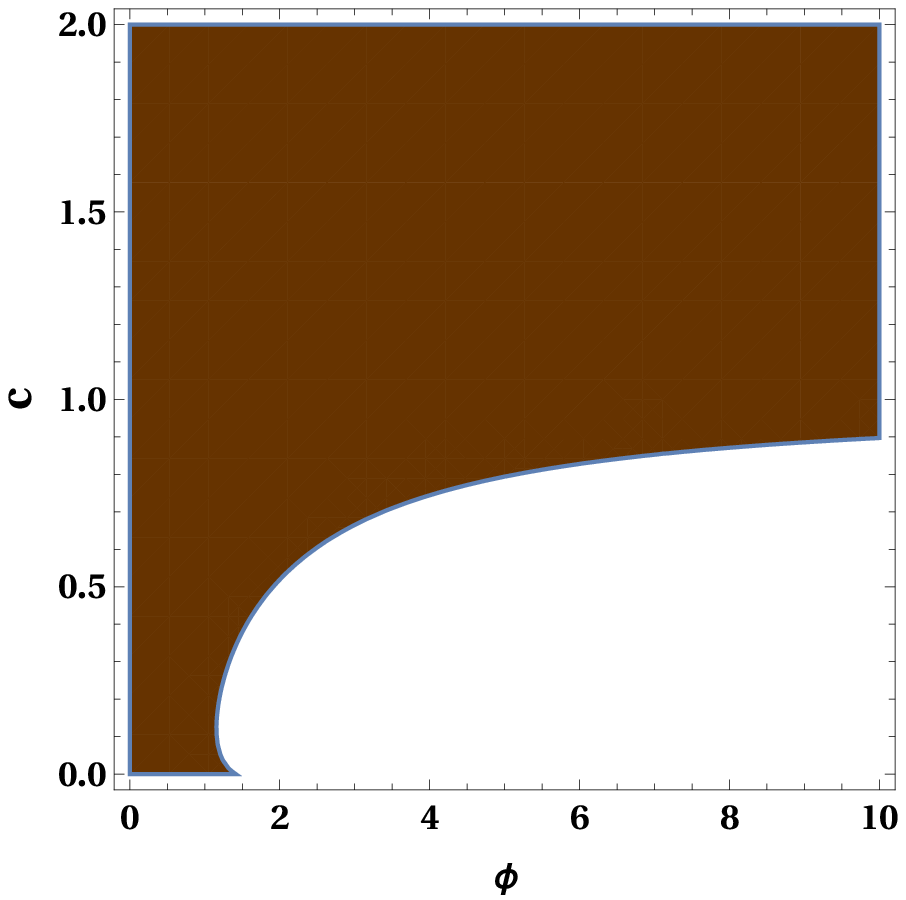}\label{fig:lam1m1}}~~
	\subfigure[$\al=4$ and $\lam=1$.]{\includegraphics[scale=.55]{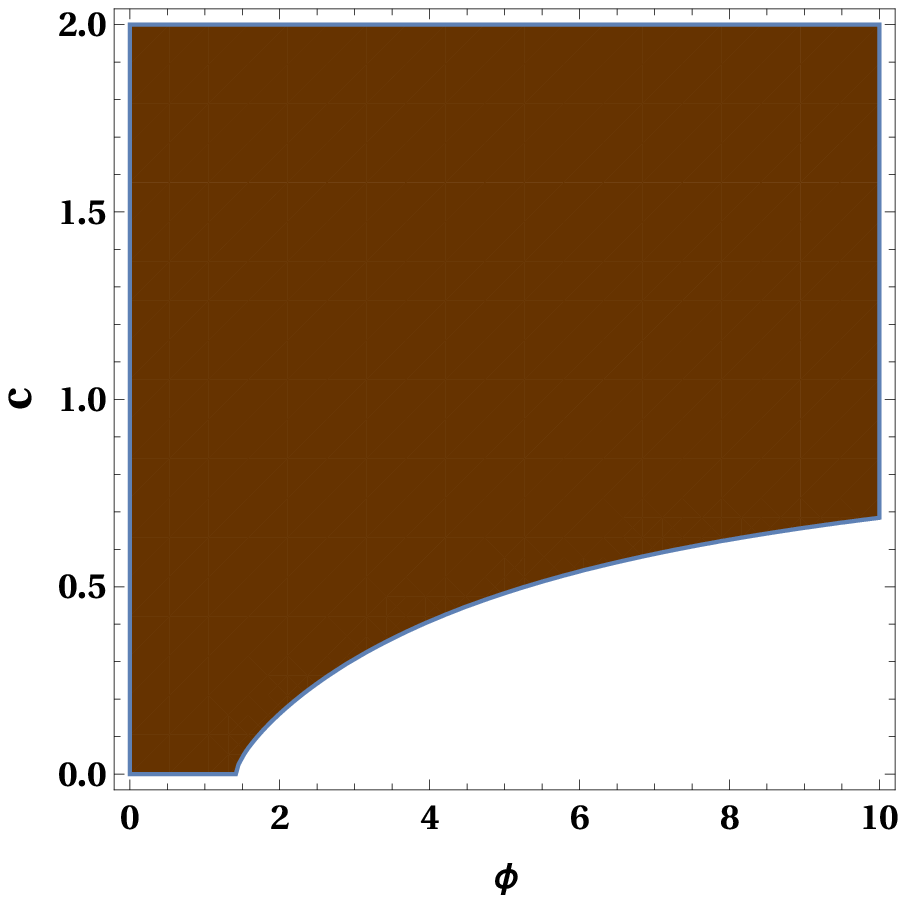}\label{fig:lam1m4}}~~
	\subfigure[$\al=10$ and $\lam=1$.]{\includegraphics[scale=.55]{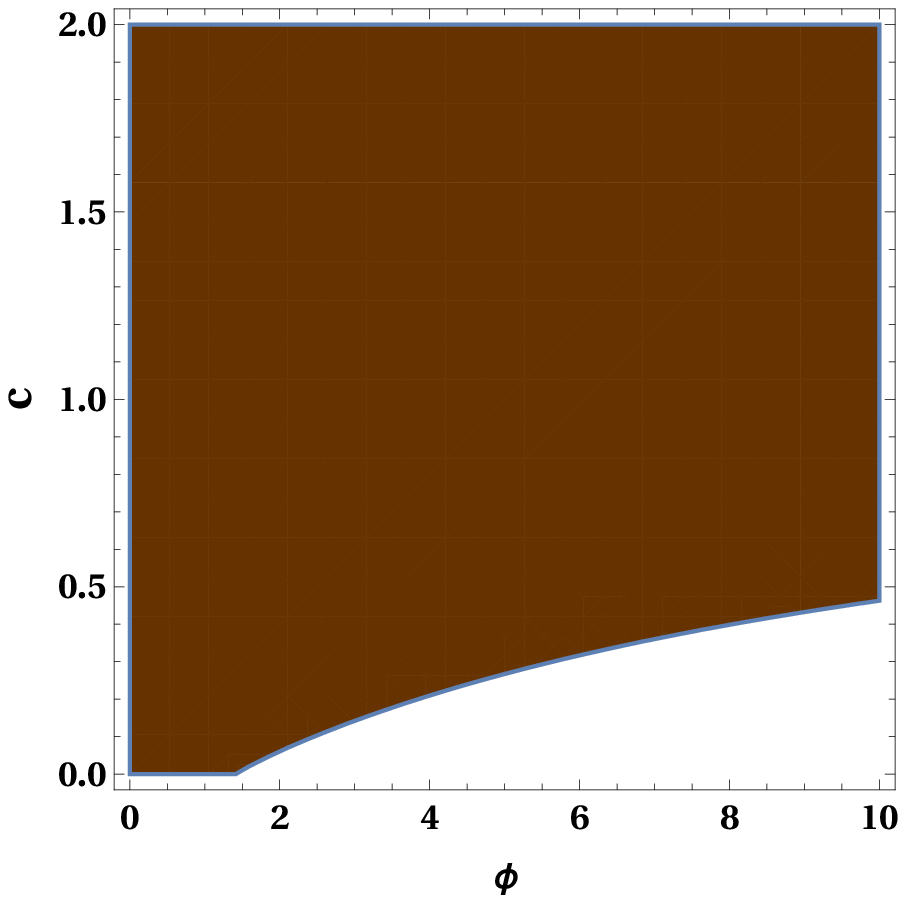}\label{fig:lam1m10}}
	\caption{Shaded regions, in the parameter space of $\phi$ and $c$, are the regions where the sWGC is true. Top figures are plotted for different values of $\lam$ while $\al$ is fixed at 6. Bottom figures are plotted for different values of $\al$ while the $\lam$ is fixed at 1. $\phi$ is in Planck mass unit.} 
	\label{fig:mlam}
\end{figure}

In order to see whether the sWGC is satisfied, given the distance conjecture and the Bousso bound, we plot the $\chi\ge 0$ as follows. Fig.~\ref{fig:mlam} shows the parameter space where $\chi(\phi)\geq 0$, {\it i.e.} the region where the sWGC (Eq.~(\ref{SWGC})) is valid in the $\phi-c$ plane. Figs.~\ref{fig:m6lam01}, \ref{fig:m6lam1} and \ref{fig:m6lam2} are plotted for three different values of $\lam$ but fixed $\al$. From these figures, we see that as we increase the values of $\lam$ the allowed regions keep getting smaller. But if we fix the value of $\lam$ and increase the value of $\al$, then we get the opposite behaviour, {\it i.e.} the allowed regions keep increasing. Figs.~\ref{fig:lam1m1}, \ref{fig:lam1m4} and \ref{fig:lam1m10} show the later case. Note that for $\phi\leq 1$, all values of the parameter $c$ are allowed {\it i.e.}, sWGC is perfectly compatible with swampland distance conjecture. In other words, the regions where the sWGC is violated appear only when $\phi\gtrsim 1.5$ and for smaller values of $c<0.5$, a region in which the distance conjecture says that towers of light states should appear. Beyond $\phi \sim 1$ (in our plots, $\phi$ has been plotted in Planck units), the distance conjecture implies that one cannot consider an EFT of the scalar field alone and other light states (typically, corresponding to extended objects) must be taken into account. From the plots, we can see that for even higher values of $\phi$, the regions where the sWGC is violated keep increasing. Thus, if the sWGC was to be true for \textit{any} value of the field, this can give us a way to put a lower bound on the value of $c$ which was assumed to be $\mathcal{O}(1)$ in \cite{Ooguri:2018wrx}\footnote{We thank the referee for pointing this to us.}. Of course, in order to systematically to do so, we need to have a better theoretical handle on the probable analytic form of the function $n(\phi)$ and we shall pursue this in future work. 

Our argument above was to motivate that the sWGC, as defined in \eqref{SWGC}, follows naturally from the swampland distance conjecture for an accelerating background. However, since we had to use the covariant entropy bound for our proof, we had to rely on some semiclassical notions of entropy of quasi-dS space which, in turn, can only be made when there is a concept of a well-defined apparent horizon. If the instabilities in the potential are so large that quantum fluctuations of the scalar disrupt the background solutions, then no such argument can be made. This is precisely the statement of the refinement in the dS conjecture. If the second derivative of the potential are so large that its magnitude is greater than that of the potential, then the tachyonic instabilities corresponding to this, on horizon scales, would spoil the definition of an apparent horizon. In other words, if we have $V''/ V < - \tilde{c}$, then our argument for the sWGC need not hold anymore. An interesting point to note here is that the sWGC, by itself, states that there can be no massive scalars, coupled to gravity, without higher self-interactions terms. 

We emphasize that our previous argument follows after the crucial assumption of having an accelerated background. This point was not so essential in the proof of the refined dS conjecture, since wherever the assumption $|V'|\leq \sqrt{2}\,V$ fails, \textit{i.e.} we have the absence of quasi-dS space, the dS conjecture would be manifestly true in those cases \cite{Ooguri:2018wrx}. In our case, however, we cannot comment much regarding the validity of the sWGC in the regime when the background is not quasi-dS. As mentioned above, this is because our argument relies crucially on the semi-classical notion of entropy in an accelerating background. Of course, this does not mean that the sWGC is not true elsewhere. This is a limitation of our argument rather than a shortcoming of the sWGC, by itself. We certainly hope that our argument can be generalized beyond accelerating backgrounds in the future in order to derive the sWGC from the swampland distance conjecture. Nevertheless, it is important to mention that quasi-dS backgrounds are typically the most interesting ones for the application of these conjectures in connection to phenomenology, such as the very early universe or even late-time acceleration. Thus, it is important to show a clean derivation of the sWGC, starting from the distance conjecture, for such backgrounds. 

Finally, note that in \cite{Gonzalo:2019gjp}, it was emphasized that \eqref{SWGC} is a \textit{local} constraint which must be satisfied for any value of the field and which applies for any scalar in the theory and not for a particular set of WGC states. This is one of the main advantages of this sWGC over previous incarnations of the sWGC. Specifically, a previous version of the sWGC \cite{Palti:2017elp} (see also \cite{Lust:2017wrl,Lee:2018spm}) required a WGC particle with mass $m(\phi)$, coupled to a light scalar $\phi$, such that in the limit of $\phi$ becoming massless, one has the relation
\begin{eqnarray}
	\partial_\phi m > m\,.
\end{eqnarray}
In this case, the presence of the WGC particle makes sure that there exists a some interactions of the scalar field which is stronger than that due to gravitational interaction. It would be nice to be able to also motivate this version of the sWGC coming from the swampland distance conjecture as well. However, for our purposes, the main roadblock that this is applicable only to WGC scalars, and not to any scalar minimally coupled to gravity, seems difficult to overcome.

\section{Cosmological consequences of the sWGC}\label{InfSec}
Let us briefly discuss a few implications of the sWGC when applied to model-building for inflation. It was stated in \cite{Gonzalo:2019gjp} that plateau-like models such as the Starobinsky potential \cite{Starobinsky:1980te} are incompatible with the sWGC and would require modification at large trans-Planckian distances. However, we point out an error in their conclusion since the authors assumed that models satisfying the sWGC must satisfy the constraint
\begin{eqnarray}
	2 \left(\frac{V'''}{V''}\right)^2 - \frac{V''''}{V''} \ge 1\,.
\end{eqnarray}
This condition, of course, follows from \eqref{SWGC}, only when $V'' >0$. This subtle point was crucially overlooked in \cite{Gonzalo:2019gjp}, which led the authors to conclude that for the Starobinsky model, the above quantity is less than $1$ beyond certain field range. We correct this misconception in the Fig.~\eqref{fig:inf} below, where we plot $\chi$, as defined in \eqref{Chi}, for the Starobinsky model. Although it is true that the model does violate the sWGC, it does so only for a very small range of field values, and more importantly in a region almost beyond the slow-roll part of the potential. 

As a consequence of the above-mentioned error, it was also concluded in \cite{Gonzalo:2019gjp} that linear potentials are naturally favored in inflationary model-building as they are the only ones left unconstrained by the sWGC. Moreover, taking  Starobinsky and Higgs inflation as stereotypical examples of plateau-like models, which are currently best fit to explain observational data, it would seem natural to argue that such models would generically face severe tension with the sWGC. However, as explained above, the conflict between such plateau-like models are minimal with the sWGC and generalizations of them can easily be made to evade this conjecture for most part of the required parameter space.

\begin{figure}[t]
	\centering
	\subfigure[$p=1/2$.]{\includegraphics[scale=.55]{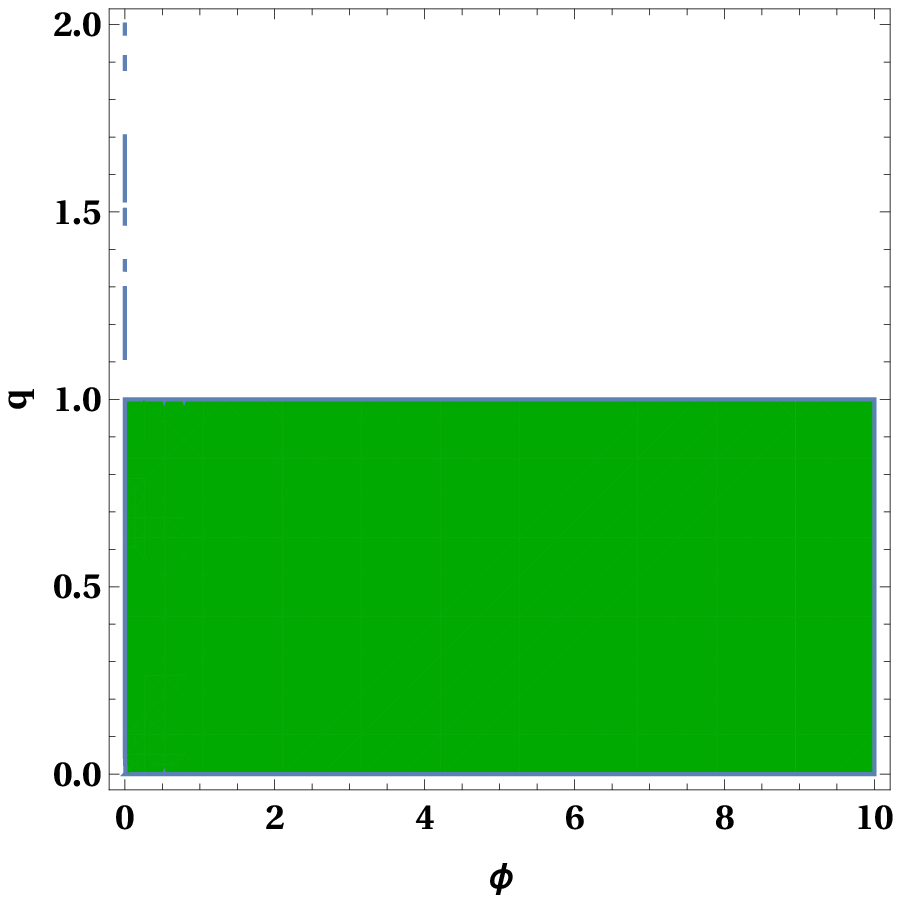}\label{fig:phalf}}~~
	\subfigure[$p=1$. The black line represents Starobinsky potential ($q=\sqrt{2/3}$).]{\includegraphics[scale=.55]{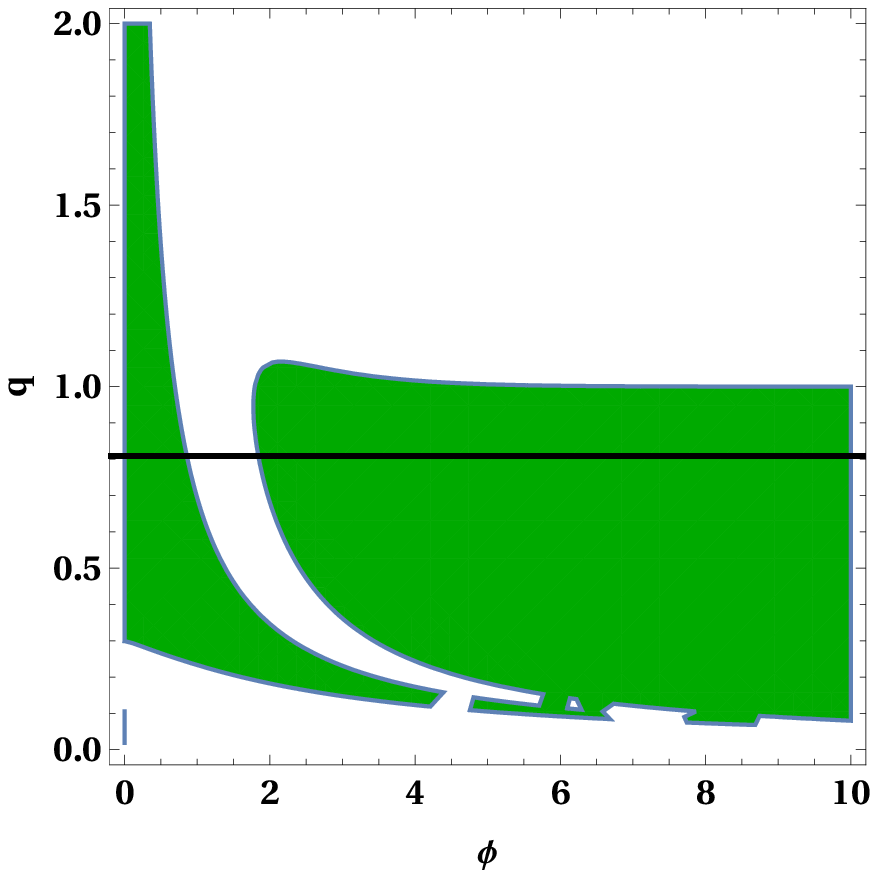}\label{fig:p1}}~~
	\subfigure[$p=10$.]{\includegraphics[scale=.55]{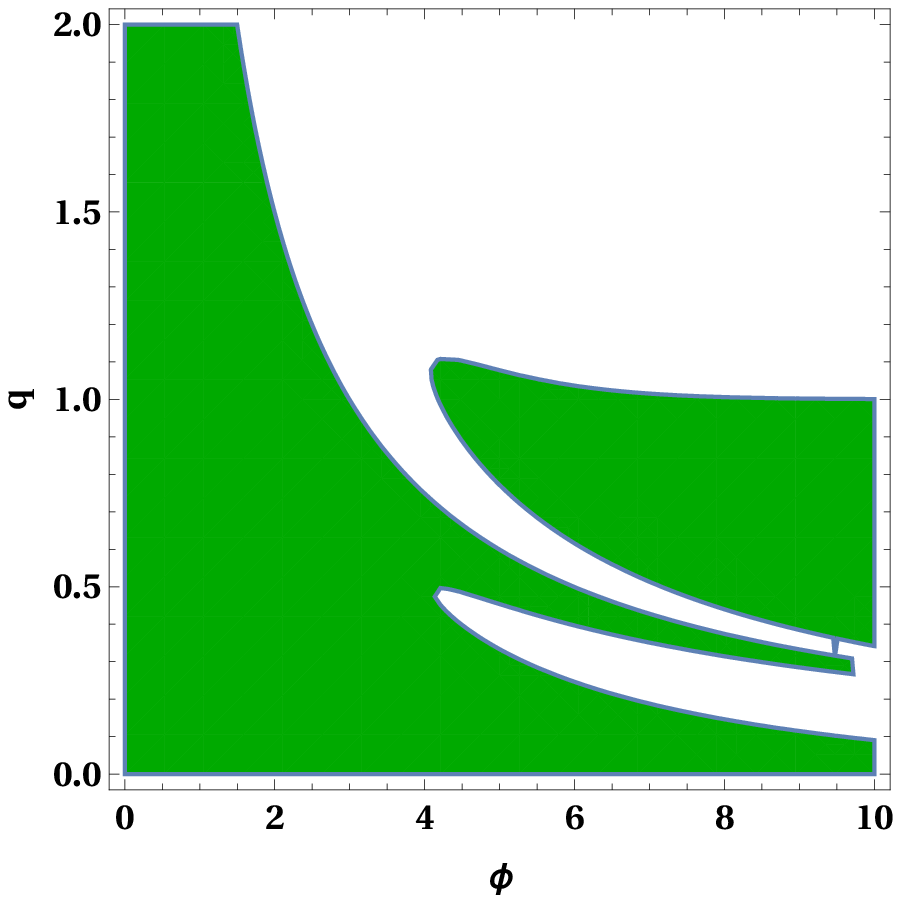}\label{fig:p10}}
	\caption{Shaded regions, in the parameter space of $\phi$ and $q$, are the regions where the sWGC is true. $\phi$ is in Planck mass unit.} 
	\label{fig:inf}
\end{figure}

To explicitly show that plateau-like models are not severely in conflict with the sWGC, let us consider a general potential of the form
\begin{eqnarray}\label{Plateau:Ex1}
	V(\phi) = V_0 (1 - e^{-q\phi})^{2p}\,,
	\label{eq:pot}
\end{eqnarray}
which never violate \eqref{SWGC} for $p=1/2$ and $q \leq 1$. The plots below explicitly proves our claims (although a quick analytic calculation for $q-1$ and $p=1/2$ serves as an example for a model which obeys the sWGC and is plateau-like). 

Fig.~\ref{fig:inf} shows the allowed parameter space where sWGC is valid in the $(\phi-q)$ space for different values of $p$. From Fig.~\ref{fig:phalf} we can see that for $p=1/2$ sWGC is valid everywhere for $q\le 1$. As can be seen for  Starobinsky-like potentials ($p=1$, Fig.~\ref{fig:p1}) there is a patch where the sWGC is not valid and this patch disconnects two allowed regions. Specifically for the Starobinsky case (black line in the Fig.~\ref{fig:p1}), a small patch between $\phi\sim 1$ and $\phi\sim 2$ is disallowed. This behaviour remains qualitatively the same for any higher values of $p$ as shown in Fig.~\ref{fig:p10}.

Of course, the refined-dS conjecture \textit{does} rule out these plateau-like models rather more stringently. However, the dS conjecture also puts bounds on super-Planckian field excursions on the linear potential and thus, we can safely conclude that the sWGC alone does not constrain the field space of inflationary model-building much. Implementing the other swampland conjectures would, however, be a much more tighter constraint on all models of inflation, as has already been noted in, say, \cite{Kinney:2018nny}. 

\section{Discussion}
One of the major thrusts in recent times has been to identify `swampland conjectures' which would enable one to categorize consistent EFTs. In the absence of a direct proof of these conjectures from first principles, it would be useful to see how many of them can be derived, starting from some of the others. One of the most tested such constraints is the swampland distance conjecture and, indeed, it has been shown that the so-called `refined dS conjecture' follows from it, when the covariant entropy bound is taken into account. In this work, we adopted a similar philosophy of applying the distance conjecture and Bousso's entropy bound, which is applicable for a quasi-dS background in the absence of large tachyonic instabilities, to show that a recently proposed sWGC follows from it. In some sense, this is a remarkable result -- we show that, at least for an accelerating universe, the fact that gravity is the weakest force naturally emerges from the distance conjecture. We end our analysis by discussing the major assumption of our argument and then a path towards a straightforward generalization of our results.

\subsection{Number of towers of light states}
The crucial assumption in our argument is the functional dependence of $n=n(\phi)$. Going beyond the obvious assumption that $n$ increases monotonically in the limit $\Delta\phi \rightarrow \Mpl$, we need to know more about the precise behaviour of higher-derivatives of this function to complete our proof. In this work, we have assumed a polynomial dependence for $n\sim \phi^\al$, with $\al\ge 1$, which seems like a natural choice to us. However, to make our results more concrete, we need to calculate the definite functional dependence of $n$, which can be done for specific cases such as in weak coupling string regimes or large radius limits\footnote{We thank Cumrun Vafa for conveying this point to us.}. Once we have the explicit behaviour of $n$ for such stringy examples, we can plug that in our estimates for the potential and its derivatives \eqref{V1}-\eqref{V4} to rigorously check if the sWGC can be derived from the distance conjecture and the covariant entropy bound. 

\begin{figure}[t]
	\centering
	\subfigure[$\al=0.1$ and $\lam=1$.]{\includegraphics[scale=.55]{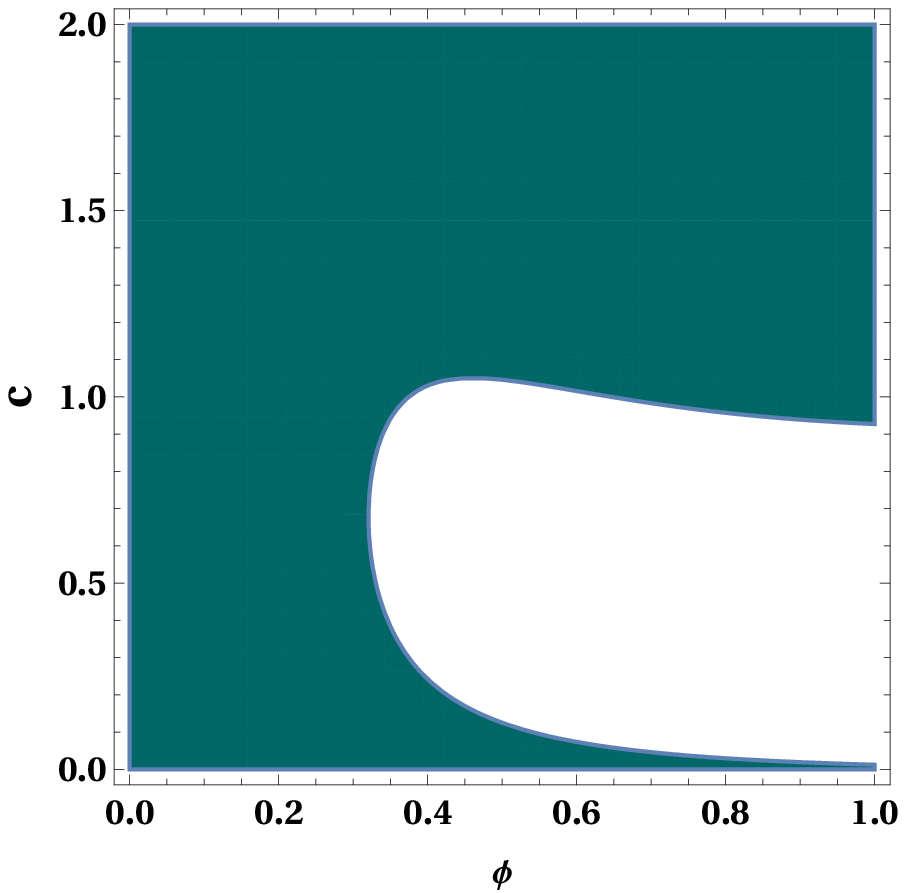}\label{fig:al_p1}}~~
	\subfigure[$\al=0.4$ and $\lam=1$.]{\includegraphics[scale=.55]{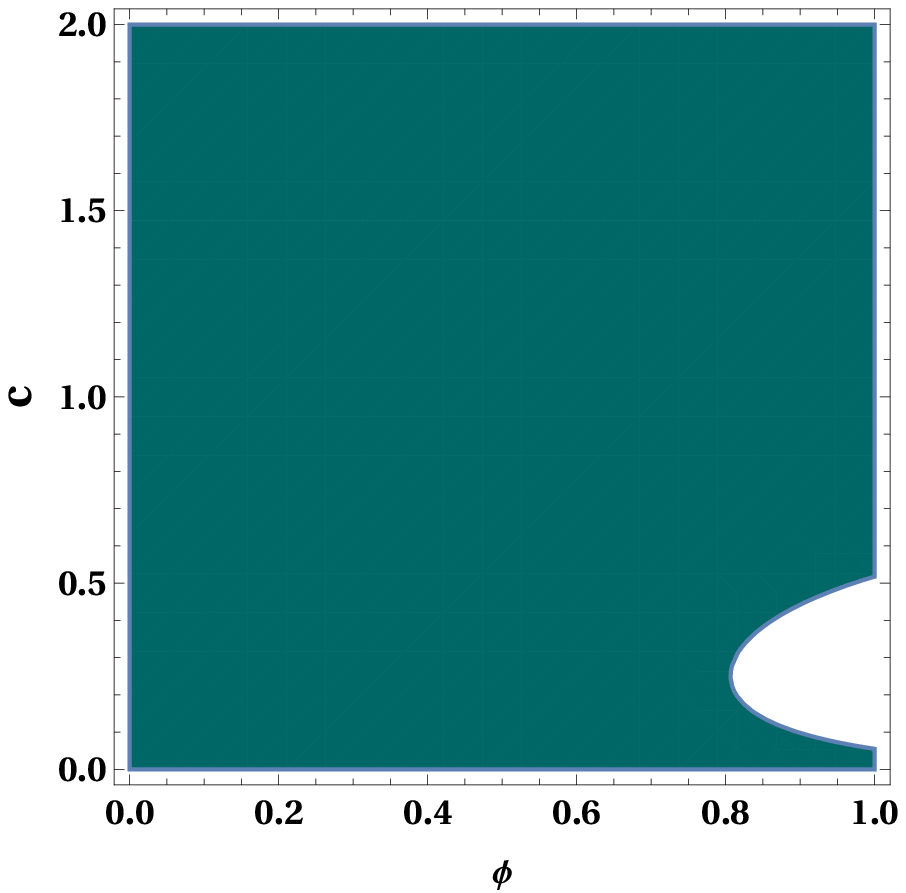}\label{fig:al_p4}}~~
	\subfigure[$\al=0.7$ and $\lam=1$.]{\includegraphics[scale=.55]{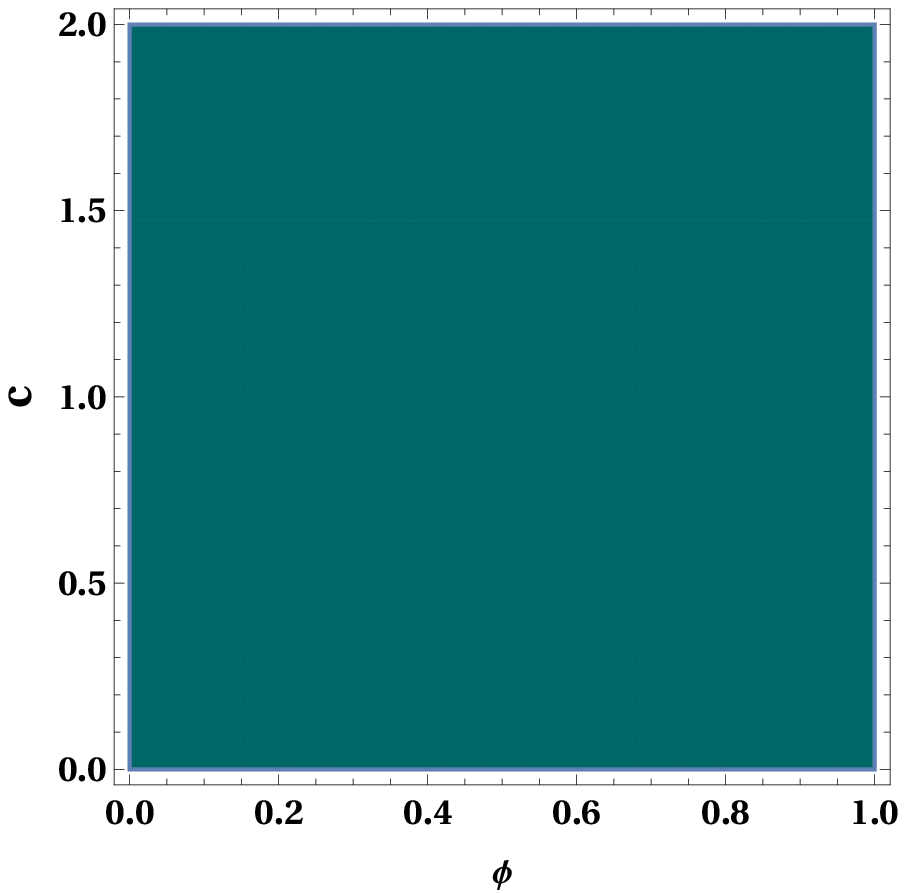}\label{fig:al_p7}} \\
	\subfigure[$\al=0.1$ and $c=1$.]{\includegraphics[scale=.55]{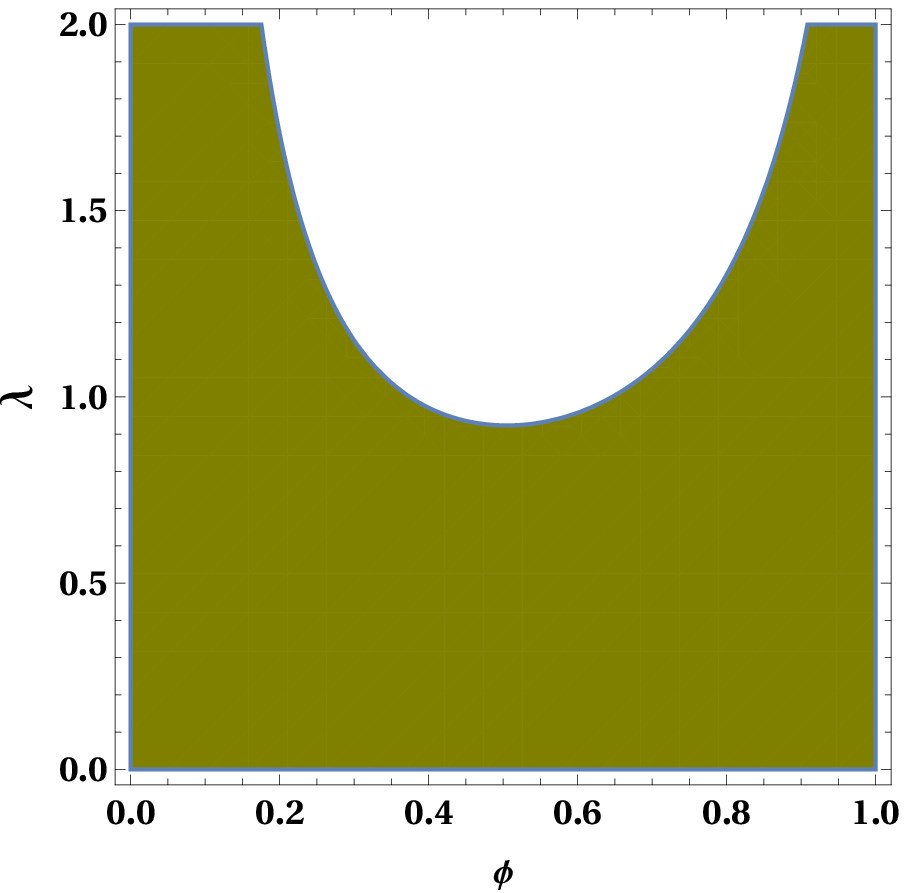}\label{fig:lam_p1}}~~
	\subfigure[$\al=0.2$ and $c=1$.]{\includegraphics[scale=.55]{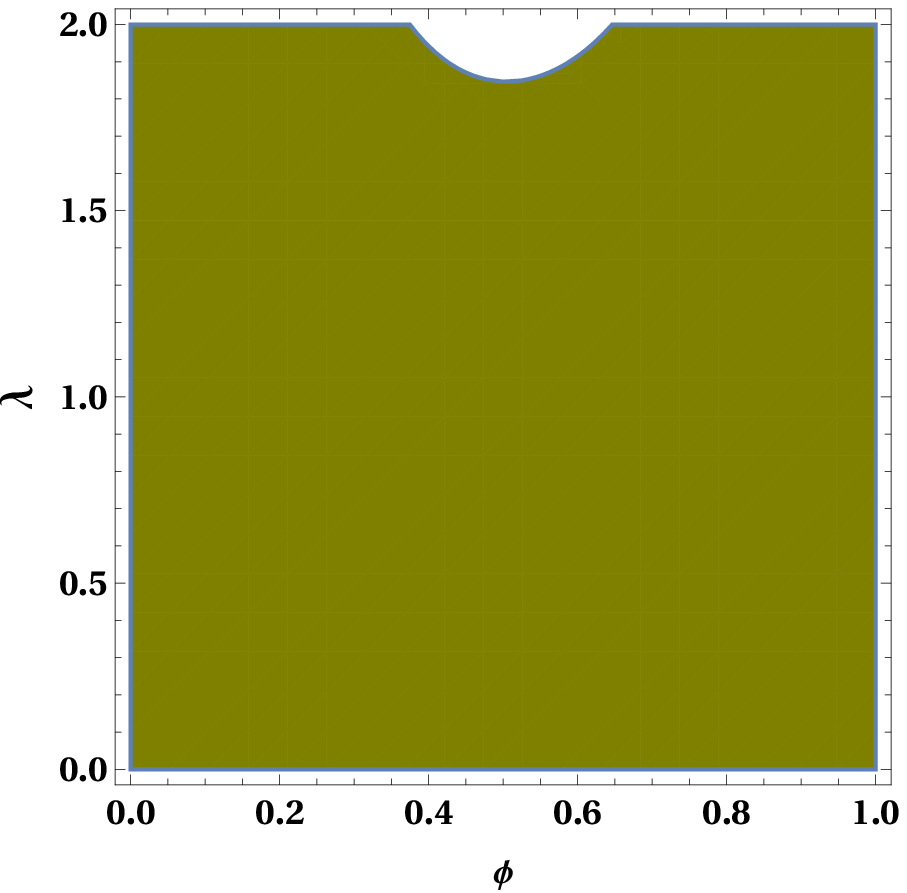}\label{fig:lam_p2}}~~
	\subfigure[$\al=0.3$ and $c=1$.]{\includegraphics[scale=.55]{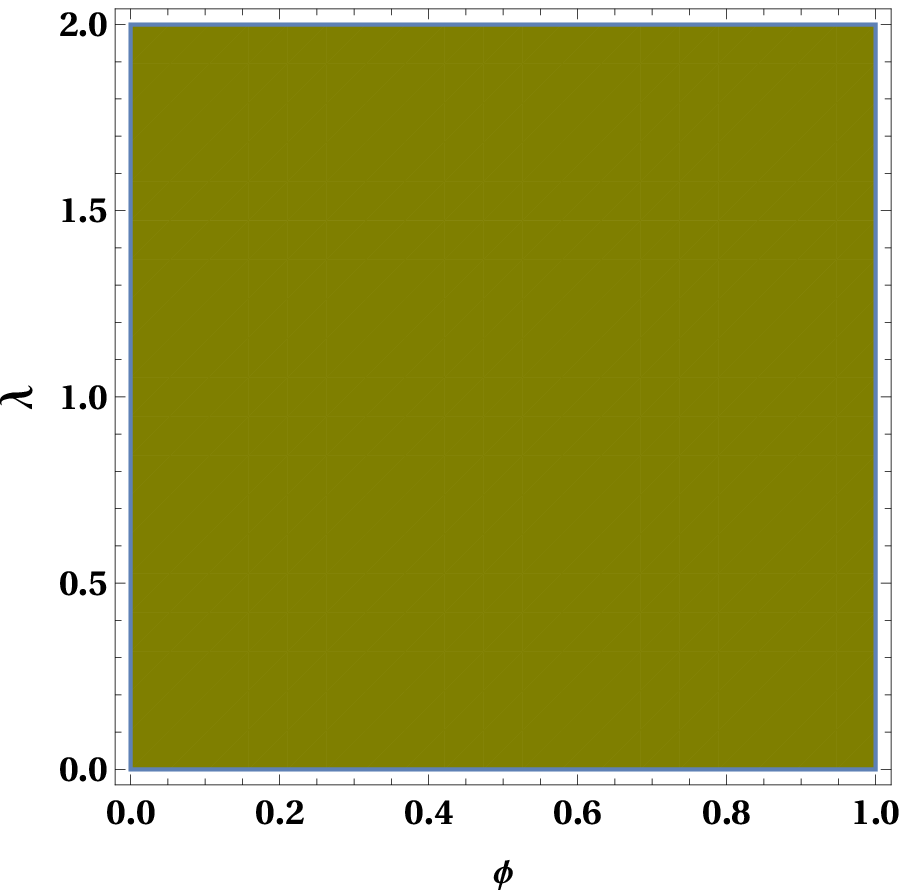}\label{fig:lam_p3}}
	\caption{Shaded regions are the allowed regions for $n\sim\phi^\al$.} 
	\label{fig:violate1}
\end{figure}

\begin{figure}[t]
	\centering
	\subfigure[$n\sim \sin(\phi\pi/2)$ and $\lam=0.1$]{\includegraphics[scale=.55]{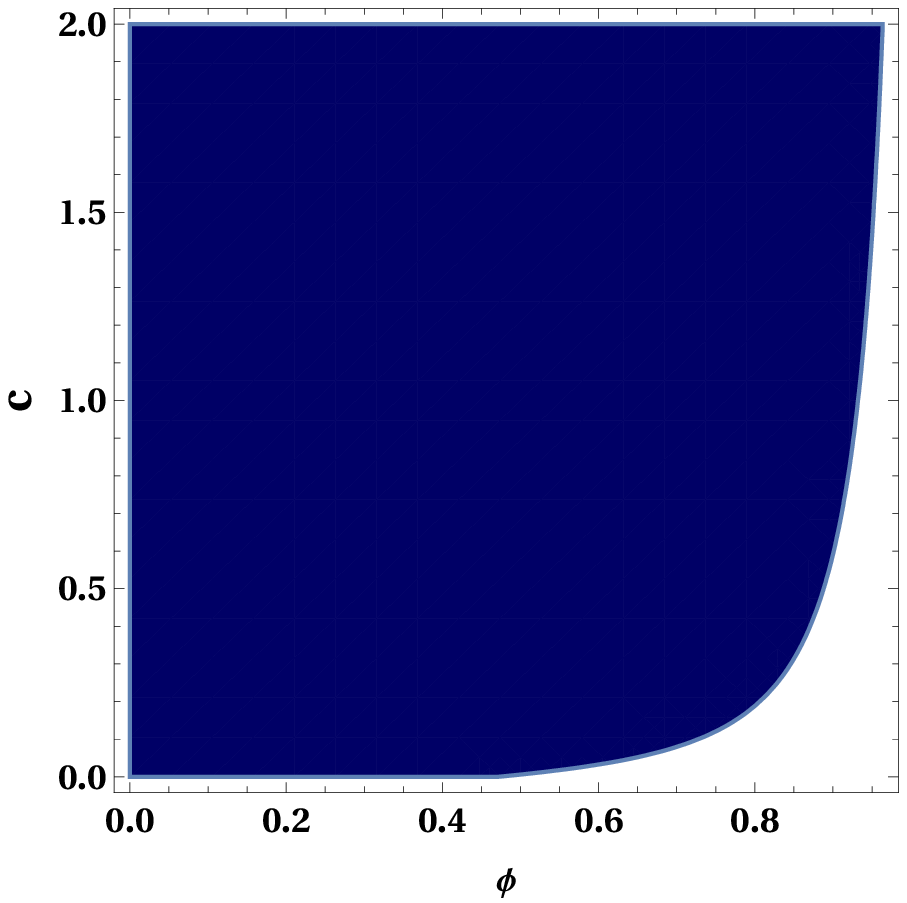}\label{fig:sin_p1}}~~~
	\subfigure[$n\sim \sin(\phi\pi/2)$ and $\lam=1$]{\includegraphics[scale=.55]{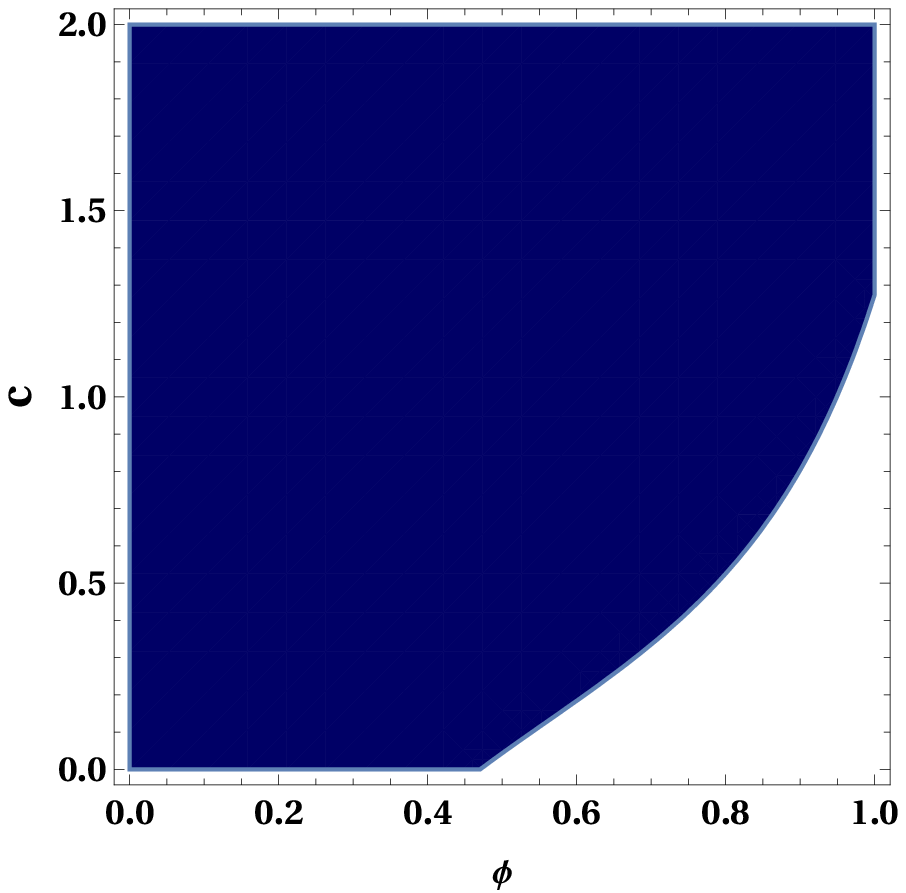}\label{fig:sin}}~~~
	\caption{Shaded regions are the allowed regions for $n\sim\sin(\phi\pi/2)$.} 
	\label{fig:violate2}
\end{figure}

In the absence of any calculation into the exact form of $n$, we have also gone on to check which kind of functions of $n$ would violate the sWGC, \textit{i.e.} lead us to have $\chi<0$ for some $\mathcal{O}(1)$ numbers $(c, \lambda)$. Keeping in mind the restriction that $n$ has to be a monotonically increasing function, we find that for specific choices such that $n$ is a concave function of $\phi$, the sWGC can be violated as the curvature of $n$ becomes more and more negative, as shown in the Figs.~\ref{fig:violate1} and \ref{fig:violate2}. This can be modeled by functional dependence of the form $n \sim \phi^\al$, with $\al\ll 1$, or with $n\sim \sin(\phi\,\pi/2)$, near $\phi \sim 1$. As shall be illustrated in the figures below, violations typically happen are for smaller values of the $c$ and larger values of $\lam$ (this is in keeping with the expectation that we reach the weakly-coupled- regime faster as $\Delta\phi\rightarrow\Mpl$), while requiring $n$ to have large negative curvature. 

From Figs.~\ref{fig:al_p1}, \ref{fig:al_p4} and \ref{fig:al_p7} we can see that the sWGC is violated for all values of $c<1.2$ but the allowed regions increase as we increase the value of $\al$. Figs.~\ref{fig:lam_p1}, \ref{fig:lam_p2} and \ref{fig:lam_p3} also show the same nature but in the ($\phi-\lam$) plane. All of these plots reflect that the regions where the sWGC holds increase with lower values of $\lam$ and higher values of $c$, still staying within the restriction that these are $\mathcal{O}(1)$ numbers, while making $\al$ smaller, \textit{i.e.} making $n$ more and more concave, increases the regions where the sWGC fails. Violation of the sWGC is also present for $n\sim \sin(\phi\pi/2)$ (Fig.~\ref{fig:violate2}). Figs.~\ref{fig:sin_p1} and \ref{fig:sin}, like Figs.~\ref{fig:lam_p1}, \ref{fig:lam_p2} and \ref{fig:lam_p3}, also show that the lower values of $\lam$ are more allowed (resulting in larger allowed range of field excursions). The bottom line from Figs.~\ref{fig:violate1} and \ref{fig:violate2}  is that, for some highly concave functions, sWGC can be violated or, in other words, our derivation for the sWGC would fail to hold starting from the distance conjecture for such choices. Note here that similar observations have already been made earlier in the literature on why it might be possible to rule out a concave nature of $n(\phi)$ \cite{Andriot:2018mav}\footnote{For an opposite point of view, see \cite{Hebecker:2018vxz}.}. In any case, it would be interesting to see if there were other constraints on the shape of $n$ coming from stringy constructions, such that there cannot be a maxima of $n$ for $\phi\le1$ or on how concave $n$ can be, which would rule out even such infractions. 

\subsection{Constraints on the shape of the potential}
It has been conjectured that coupling an EFT to quantum gravity leads to severely restricting the form of allowed potentials of the scalar field in the low-energy theory. The remarkable conclusion of \cite{Ooguri:2018wrx} was to show that one can come to this result, starting from the distance conjecture and applying the covariant entropy bound, without requiring any information regarding the difficulty of construction of dS vacua in string theory. A straightforward extension of their result was considered in this work to show that one can even arrive at the sWGC from a similar starting point. However, even allowing for the current limitations of our work as outlined in the previous section, one can nevertheless further postulate an infinite number of inequalities on the shape of the potential. To illustrate our point, consider our expression for the second derivative of the potential \eqref{V2}, in the weak coupling limit, when the Bousso bound gets saturated. Indeed, if we get a specific form of the function $n$ from some string theory example, it would be possible to place inequalities relating the second derivative of the potential to the first derivative and the potential itself. (This would be, in essence, further refining the dS swampland conjecture as in \cite{Andriot:2018mav}.) One can then continue this process iteratively for higher derivatives generally, and not specifically for the particular expression which appears in the sWGC alone. The promise of such a program would be immense -- the ability to highly constrain shapes of allowed potentials in low-energy EFTs, coming from quantum gravity, which were hitherto not present when gravity is not taken into account. This also nicely connects to our heuristic comments about similarities between the swampland program and early days of quantum theory made in the introduction. However, we leave this for future work after we are investigate the general form of the function $n$ in some explicit examples. 

Let us make a final comment regarding the shape of the potential in the least constrained case. For the dS conjecture alone, this means satisfying the equation $V' = -c\,V$, leading to an exponential form of the potential. In fact, revisiting the derivation of the refined dS conjecture as in \cite{Ooguri:2018wrx}, we find that this (steep) exponential form of the potential satisfies all the higher derivative expressions \eqref{V1}-\eqref{V4}, provided we do not have large tachyonic instabilities in the potential. In our work, we find that the exponential form of the potential remains the solution for the least constraining case (\textit{i.e.} when $n$ is a constant function). 

\section{Note added: Modified version of the sWGC}\label{Note}
While we were completing this work, an updated version of \cite{Gonzalo:2019gjp} corrected the form of the sWGC. However, the gist of our discussion in the section on its applications to inflationary potentials (Sec.~\ref{InfSec}) is still applicable to the modified sWGC as we demonstrate as follows. The modified sWGC states that 
\begin{eqnarray}\label{sWGCmod}
	\chi(\phi)_{\rm mod}:= 2 \left(V'''(\phi)\right)^2 - V''''(\phi) V''(\phi) - \left(V''(\phi)\right)^2 > 0\,.
\end{eqnarray}

\begin{figure}[t]
	\centering
	\subfigure[$p=1/2$.]{\includegraphics[scale=.55]{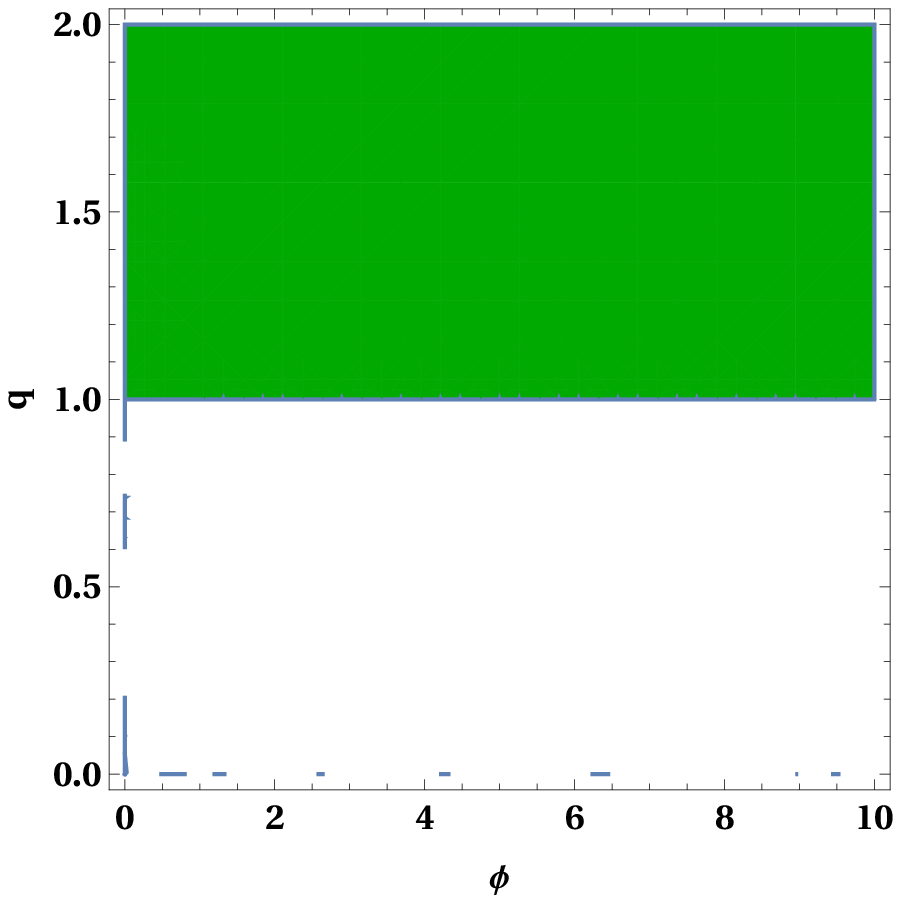}\label{fig:phalf_new}}~~
	\subfigure[$p=1$. The black line represents Starobinsky potential ($q=\sqrt{2/3}$).]{\includegraphics[scale=.55]{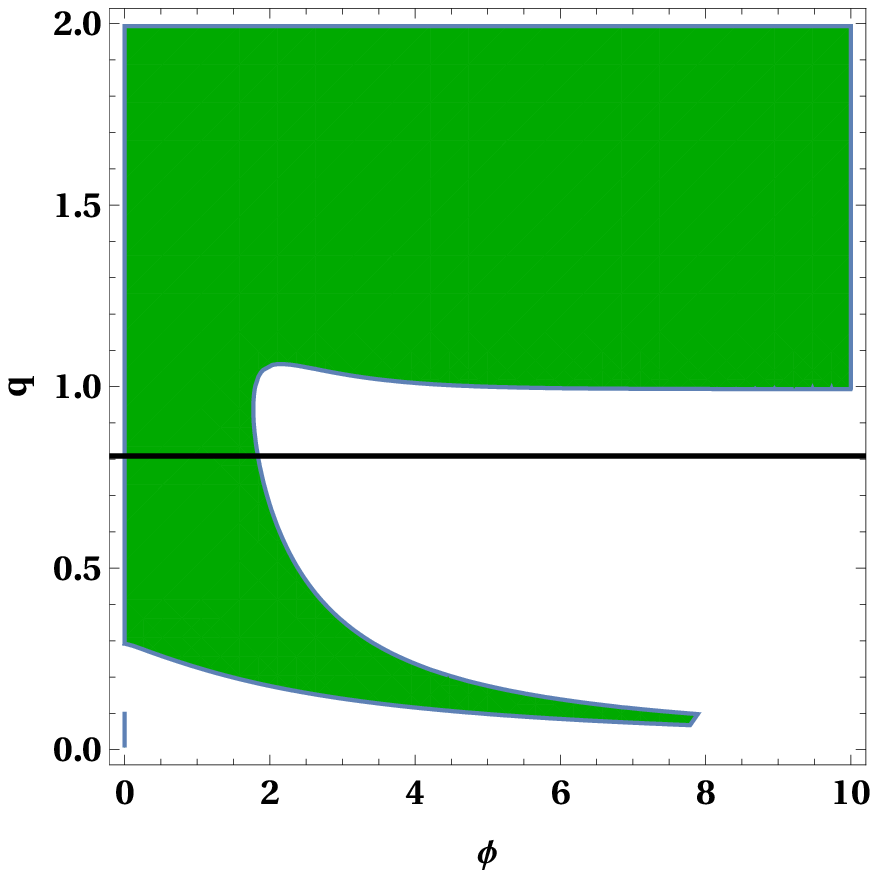}\label{fig:p1_new}}~~
	\subfigure[$p=10$.]{\includegraphics[scale=.55]{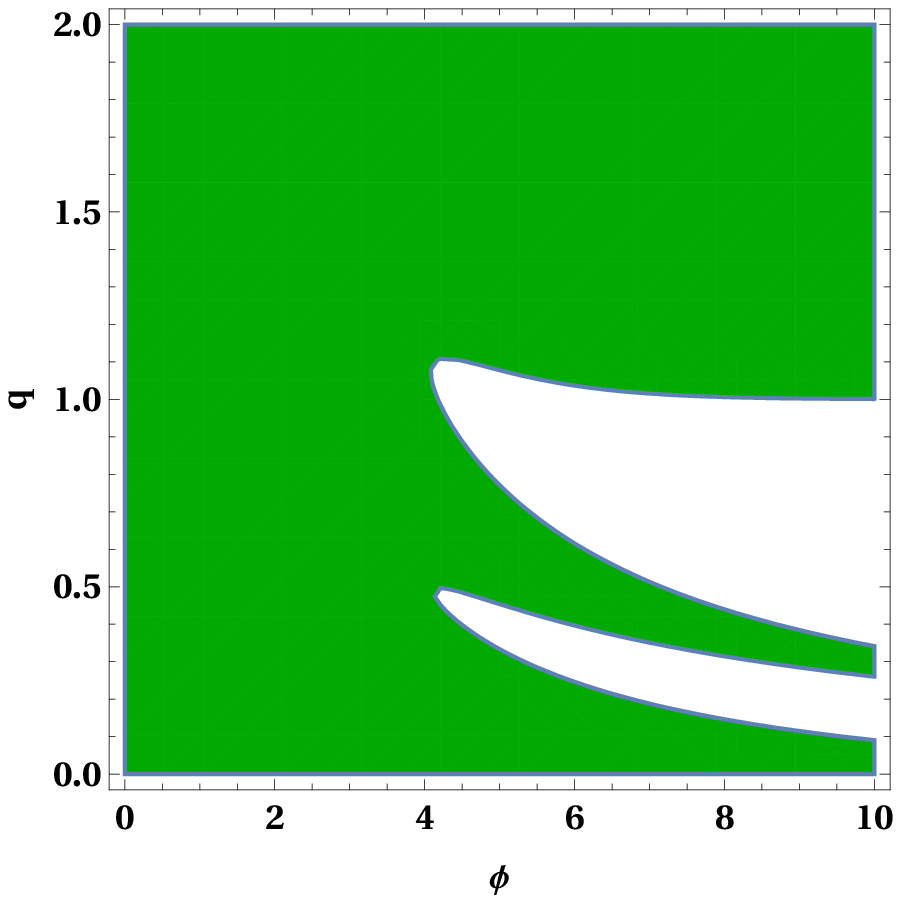}\label{fig:p10_new}}
	\caption{Shaded regions, in the parameter space of $\phi$ and $q$, are the regions where the sWGC is true. $\phi$ is in Planck mass unit. We have considered the potential~(\ref{eq:pot}).} 
	\label{fig:inf_new}
\end{figure}

If we apply our analysis for plateau-like models to plot $\chi(\phi)_{\rm mod}$, we can easily see that the discontinuities in $\chi$ are now ameliorated due to the absence of $V''$ in the denominator. However, our example of a plateau-like model (Eq.~(\ref{eq:pot})), which obeyed the original form of the sWGC \eqref{SWGC}, also obeys the modified version \eqref{sWGCmod}. Thus, our main conclusion that the sWGC is not very constraining for inflationary model-building remains valid even in this case. However, even more interestingly, we find that for the modified version of the sWGC, the allowed regions increase manifold as compared to the previous version as one can easily see by comparing the Figs.~\ref{fig:inf} and \ref{fig:inf_new}. In fact, for Starobinsky-like potentials ($p=1$), the new version is fully compatible for $q>1$ (Fig.~\ref{fig:p1_new}).

Finally, we also note that, starting from the swampland distance conjecture, we find that the modified sWGC \eqref{sWGCmod} follows exactly as the older version without any changes. This is simply because the older and newer versions only disagree when $V''<0$; however, this is not the case for the potential we need to consider for our arguments. As an aside, it is curious to note that the modified sWGC rules out any (strictly) quadratic potential (even with $m^2<0$) which means that there are no allowances for tachyonic instabilities in the potential any longer.

\vskip15pt

\section*{Acknowledgments:} \noindent  We are grateful to Cumrun Vafa and Eran Palti for helpful comments on an earlier version of this draft. This research was supported in part by the Ministry of Science, ICT \& Future Planning, Gyeongsangbuk-do and Pohang City and the National Research Foundation of Korea (Grant No.: 2018R1D1A1B07049126).

\end{document}